\documentclass[sigconf]{aamas}

\usepackage{balance} 

\usepackage{placeins}

\doi{CEJT6762}

\makeatletter
\gdef\@copyrightpermission{
  \begin{minipage}{0.2\columnwidth}
   \href{https://creativecommons.org/licenses/by/4.0/}{\includegraphics[width=0.90\textwidth]{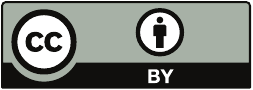}}
  \end{minipage}\hfill
  \begin{minipage}{0.8\columnwidth}
   \href{https://creativecommons.org/licenses/by/4.0/}{This work is licensed under a Creative Commons Attribution International 4.0 License.}
  \end{minipage}
  \vspace{5pt}
}
\makeatother

\setcopyright{ifaamas}
\acmConference[AAMAS '26]{Proc.\@ of the 25th International Conference
on Autonomous Agents and Multiagent Systems (AAMAS 2026)}{May 25 -- 29, 2026}
{Paphos, Cyprus}{C.~Amato, L.~Dennis, V.~Mascardi, J.~Thangarajah (eds.)}
\copyrightyear{2026}
\acmYear{2026}
\acmDOI{}
\acmPrice{}
\acmISBN{}

\title[Fairness Dynamics in Digital Economy Platforms with Biased Ratings]{Fairness Dynamics in Digital Economy Platforms with Biased Ratings}

\author{\href{https://orcid.org/0000-0001-5466-1259}{Martin Smit}}
\affiliation{
  \institution{University of Amsterdam}
  \city{Amsterdam}
  \country{Netherlands}}
\email{j.m.m.smit@uva.nl}

\author{\href{https://orcid.org/0000-0002-2310-6444}{Fernando P. Santos}}
\affiliation{
  \institution{University of Amsterdam}
  \city{Amsterdam}
  \country{Netherlands}}
\email{f.p.santos@uva.nl}

\begin{abstract}
The digital services economy consists of online platforms that facilitate interactions between service providers and consumers.
This ecosystem is characterized by short-term, often one-off, transactions between parties that have no prior familiarity.
To establish trust among users, platforms employ rating systems which allow users to report on the quality of their previous interactions.
However, while arguably crucial for these platforms to function, rating systems can perpetuate negative biases against marginalised groups.
This paper investigates how to design platforms around biased reputation systems, reducing discrimination while maintaining incentives for all service providers to offer high quality service for users.
We introduce an evolutionary game theoretical model to study how digital platforms can perpetuate or counteract rating-based discrimination.
We focus on the platforms’ decisions to promote service providers who have high reputations or who belong to a specific protected group.
Our results demonstrate a fundamental trade-off between user experience and fairness: promoting highly-rated providers benefits users, but lowers the demand for marginalised providers against which the ratings are biased.
Our results also provide evidence that intervening by tuning the demographics of the search results is a highly effective way of reducing unfairness while minimally impacting users.
Furthermore, we show that even when precise measurements on the level of rating bias affecting marginalised service providers is unavailable, there is still potential to improve upon a recommender system which ignores protected characteristics.
Altogether, our model highlights the benefits of proactive anti-discrimination design in systems where ratings are used to promote cooperative behaviour.
\end{abstract}


\newcommand{\BibTeX}{\rm B\kern-.05em{\sc i\kern-.025em b}\kern-.08em\TeX}

\begin{document}


\pagestyle{fancy}
\fancyhead{}


\maketitle 


\section{Introduction}

Evidence of discrimination against marginalised communities on online platforms is widespread in both the academic community and beyond.
Research has found that guests on Airbnb with Black-sounding names are accepted 16\% less often than guests with White-sounding names~\cite{edelman_racial_2017} and
Black Airbnb hosts in the United States charge 5-7\% less than White hosts for equivalent properties~\cite{jaeger_racial_2023}.
On Uber, users with Black-sounding names experience twice the cancellation rate than White-sounding names~\cite{ge_racial_2020}.
On eBay, after accounting for reviews, it was found that listings with photos showing a Black hand holding a baseball card sold for 20\% less on average than when a White hand was photographed~\cite{ayres_race_2015}, and that women receive fewer and lower bids than men when selling identical items in new condition, leading to, again, a final sell price of 20\% less on average~\cite{kricheli-katz_how_2016}.
Such biases do not require explicit group identifiers. Evidence shows that when information about a user or service provider's demographic characteristics are not directly available, they can be inferred~\cite{kricheli-katz_how_2016}.
Furthermore, other information can be used as a proxy, such as on Airbnb where the neighbourhood majority ethnicity is a statistically significant predictor of price after controlling for all observable features~\cite{goel_tackling_2020}.

While research indicates that discrimination based on demographic characteristics can be partially alleviated by rating systems~\cite{abrahao_reputation_2017}, these systems have, themselves, been shown to be susceptible to bias~\cite{goel_tackling_2020, hannak_bias_2017} leading to lower demand, prices, and revenue for those discriminated against~\cite{tuebner_price_2017}.
As argued by \citet{van_doorn_migration_2023}, while precise, cross-platform data on the socioeconomic, migratory, and residence status of those working on digital platforms is unavailable, 
data does indicate that migrant workers are hugely overrepresented on these platforms, even in countries with large domestic labour markets such as China and India \cite{chen_digital_2019, tandon_sustaining_2024, raval_platform-living_2020}.

Thus, a large number of economically (and often legally) vulnerable individuals are impacted by bias in the local populations for which they provide services.
For many, this bias affects a significant proportion of their total earnings, as data from a 2022 Eurostat survey revealed that platform labour made up more than 75\% of total income for almost a 25\% of participants who reported using digital platforms for employment~\cite{noauthor_employment_2023}.

Given the importance placed on digital economies by various (inter-)governmental bodies such as the US Bureau of Labor Statistics~\cite{dell_assessing_2020} and the European Commission~\cite{noauthor_directive_2024}, there is a rich body of research analysing these problems from a legal, policy, and econometric view.
Although the potential sources of bias have been extensively identified, and many potential solutions proposed, there are few predictive models that explain the circumstances and mechanisms leading to unfair outcomes ~\cite{monachou_discrimination_2019, che_statistical_2024}.
Indeed, as highlighted in~\cite{monachou_discrimination_2019} as an open problem, it is unclear the extent to which the decisions made by the platform designers can perpetuate or alleviate the biases of the platform's users.
Furthermore, besides one suggestion we discuss in Section~\ref{sec:related-models-of-dse}, these works stop short of making explicit recommendations to platform designers on how to improve the situation.

In this paper we address this issue by analysing the role of the recommendation algorithm responsible for facilitating matches between users and providers on digital platforms.
We introduce an evolutionary game-theoretical model of a digital platform in which providers adapt to user- and platform-based incentives to maintain high quality service, a trait essential to the longevity of a platform.
This allows us to identify the emergent strategic and utility effects of platform designers' decisions to tweak their algorithms by showing more or fewer providers of a certain rating or group.

Our contributions can be summarised as follows:
\begin{enumerate}
    \item We formulate an evolutionary game theory (\textbf{EGT}) model that captures how different groups of agents respond to and are affected by recommender systems.
    \item We numerically analyse the model's strategy dynamics to define an intuitive necessary condition for a platform to sufficiently incentivise high-effort behaviour.
    \item Using the model, we highlight the trade-off between user experience and group-level fairness faced by platform designers under the constraint of treating all providers equally despite rating bias.
    \item We demonstrate how this trade-off vanishes when the group against which the population is biased can be algorithmically prioritised and that there is still a benefit in doing so under moderate uncertainty in the extent of the bias.
\end{enumerate}

\paragraph{Structure of the paper}
In Section~\ref{sec:related-lit}, we summarise the related literature on fairness in two-sided recommender systems, evolutionary models of fairness more generally, and the two aforementioned mathematical models of bias in online platforms.
In Section~\ref{sec:model} we formulate the problem and introduce our model and its dynamics.
Then, we present the experimental design of our simulations in Section~\ref{sec:experiments} and the evaluation metrics we use to judge the effectiveness of platforms in Section~\ref{sec:evaluation-metrics}.
In Section~\ref{sec:res-strategy-dynamics} we analyse how the model's dynamics respond to different parameter setups and define the necessary condition for a platform to succeed based on its strategy dynamics.
Subsequently, we show in Section~\ref{sec:pareto-front} how forcing algorithmic equal treatment creates a trade-off between provider fairness and user experience, and how is no longer the case when this constraint is removed in Section~\ref{sec:anti-discrimination}, and even when uncertainty is introduced in Section~\ref{sec:uncertainty}.
Finally, in Section~\ref{sec:conclusion} we discuss the implications our results have for designers wanting to improve platforms, as well as avenues for future research.
Our code (and the appendix) is available at \cite{smit_appendix_2026}.

\section{Related literature}\label{sec:related-lit}

\subsection{Fairness in two-sided recommender systems}
Underpinning digital platforms are machine learning algorithms that attempt to reduce search friction for users by showing them relevant content, items, or service providers.
A key challenge identified in such systems is the difficulty of  simultaneously providing accurate recommendations for both the consumer side which are fair for the producer side~\cite{biswas_toward_2022, patro_fair_2022}.

One fundamental difficulty discussed extensively in \cite{deldjoo_fairness_2024} is that there are many, sometimes incompatible, definitions of fairness in recommender systems, each carrying an implicit set of normative judgements about the world.
In \cite{geyik_fairness-aware_2019} and \cite{suhr_does_2021}, as well as our paper, we identify group-level fairness, as opposed to individual fairness, to be the most effective way to examine how protected characteristics can affect individuals.

\citet{geyik_fairness-aware_2019} provide specific algorithms which they show, both theoretically and in practice on LinkedIn data, are able to ensure the distribution of protected characteristics in the top $k$ search results follows some desired distribution.
\citet{suhr_does_2021} then apply one of these algorithms to explore how the interactions between fair ranking algorithms and the inherent gender biases of employers can have a real impact on hiring decisions from online platforms.
They conclude that the effectiveness of such algorithms is clear but that it may also be lessened by gender biases related to the task for which the candidate is being hired.
Building on these compartmental studies, our paper shows the conditions under which such fair algorithms can be deployed without detrimentally harming the incentive structures that encourage high-effort behaviours.
As we discuss in Section~\ref{sec:evaluation-metrics}, we measure fairness by demographic parity ratio due to the ubiquity and simplicity of parity-based metrics.

Another concept which is important for two-sided recommender systems is the temporal nature of fairness~\cite{chakraborty_fair_2017,patro_fair_2022}.
Given our use of stochastic processes to model digital platforms, our model explicitly includes an element of time, which allows us to extract both instantaneous and long-term measurements of fairness, the latter of which being the focus of this paper.



\subsection{Fairness in multi-agent (reputation) systems}
A large body of work studies how costly cooperation can be sustained in large populations through utilising reputations and \textbf{indirect reciprocity}~\cite{ohtsuki_how_2004, santos_complexity_2021}.
Recent work~\cite{smit_learning_2024} shows how poorly designed reputation systems 
can lead to unfair outcomes, even when agents differ only in some arbitrary group label.
The findings in \cite{smit_learning_2024} highlight that such biased reputation systems can prevent both fairness and universal (i.e. not based on group) cooperation. Other work considers how unfair outcomes can result from cooperation incentives and the emergence of different tasks in populations of learning agents \cite{kimfair}.
In our paper we also consider the interplay between fairness and cooperation, yet we focus on digital economy platforms and consider a recommendation and choice system more realistic than \cite{smit_learning_2024}, in which agents have an equal likelihood of interacting with anyone.

\subsection{Correcting biased ratings}
Some suggestions for handling biased ratings are summarised by~\citet{tushev_systematic_2022}.
Beyond implementing structured reputation systems which encourage objective evaluation of interactions~\cite{pallais_inefficient_2014}, 
adjusting or reinterpreting suspected biased reviews has been suggested as a way to achieve fairness in digital platforms~\cite{hannak_bias_2017}.
This was later explored by
\citet{goel_tackling_2020}, who show how platforms may be able to control for bias by performing post-aggregation bias correction which aims to minimally alter the aggregated ratings of all individuals such that the correlation between sensitive attribute and score is at most some value $\delta$.
This paper acknowledges, however, that while the proposed mechanisms work in theory, the implementation of such policies faces issues in determining the acceptable loss in information and how much bias in the ratings is permissible.
Given that rating biases may be impossible to eliminate entirely, and the line between what is considered a ``good'' and ``bad'' rating being thin, our study explores how platforms can address rating bias even when it cannot be fully corrected.

\subsection{Modelling fairness in digital platforms}\label{sec:related-models-of-dse}

Efforts have been made to develop a mechanistic understanding of the effects of bias in online platforms.
\citet{monachou_discrimination_2019} examine a digitally-mediated labour market where employers are matched with workers and must choose whether to hire or reject the candidate.
Employers leave ratings for workers they hire which cumulatively reduce the uncertainty other employers have about the type of the worker, which can be either high- or low-skilled.

The authors exogenously introduce bias through \textit{discriminating employers}, who make up a certain proportion of the market and hold the misspecified prior belief that the likelihood that a minority worker without any reviews is high-skilled is only $\beta$ times as much as majority workers with $\beta \in (0, 1)$.
Thus, even if ratings accurately reflect performance, bias in interpreting a group's ratings can affect the rate at which employers acquire accurate information about worker ability, ultimately affecting their utility.

A similar conclusion is reached by \citet{che_statistical_2024}, who discuss how \textit{statistical discrimination}, the inference of an individual's characteristics through their group's characteristics, can arise due to a self-fulfilling perception that the same numerical rating is a less reliable reflection of the underlying quality of an agent in one group than in another.
In their model, ratings gradually become inaccurate over time as agents stochastically change their ``type'' (either high- or low-quality) at rate $\delta > 0$.
Due to this, whichever group is chosen more often will have more up-to-date ratings and thus be trusted and favoured by potential users.

In their analysis, the authors of both papers mostly hold the design choices of the platform as exogenous, apart from one section in \cite{monachou_discrimination_2019} where employers who are identified as ``discriminating'' through a pattern of discriminatory rejection are not matched with minorities until uncertainty about their skill is small enough.
Even in this suggestion, however, they concede that their notion of fairness is one-sided, taking workers, but not employers, into account.
Ultimately, while these models provide a good understanding of why rating bias is so pervasive, suggestions of how to address this issue have not been studied taking into account all the stakeholders' perspectives i.e., providers, users, the platform itself.

\section{Problem Formulation}\label{sec:model}
\begin{figure}[t]
    \centering
    \includegraphics[scale=0.35]{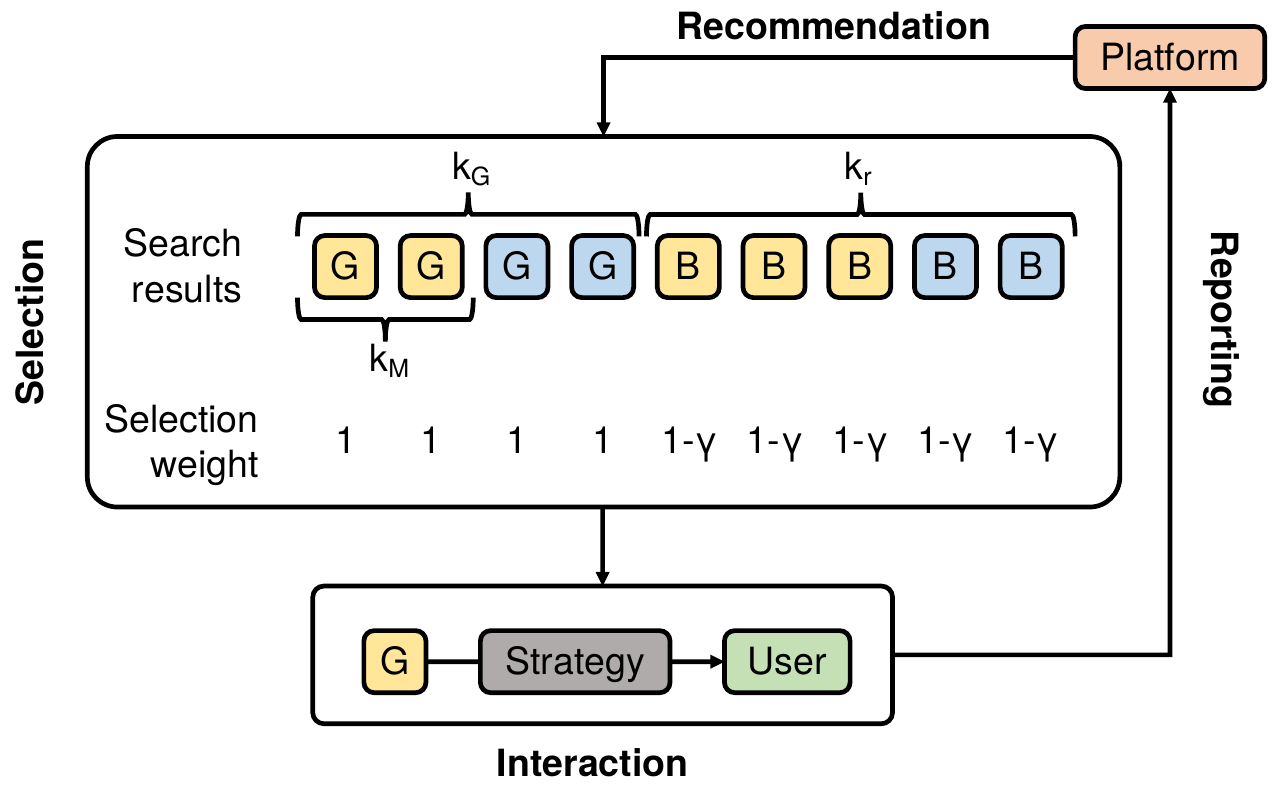}
    \caption{
        We model interactions on digital service economy platforms with four distinct steps.
        First, the platform recommends a list of $k$ providers to a user, $k_G$ of which are rated $\mathit{Good}$; out  of these, $k_M$ are part of a marginalised community.
        The user then chooses randomly from the list, giving those with a $\mathit{Good}$ rating unit weight, and a $\mathit{Bad}$ rating weight $1-\gamma$.
        The chosen provider  acts according to their strategy, either $\mathit{high}$ or $\mathit{low}$-$\mathit{effort}$, rewarding the provider $b-c$ or $b$ utility respectively.
        The user reports the action taken to the platform, who updates the rating of the provider.
        User reports are biased: with rate $\epsilon$, they will report a marginalised user who played $\mathit{High}$ as $\mathit{Bad}$ even when they should be rated $\mathit{Good}$.
    }
    \label{fig:model-diagram}
\end{figure}

We model interactions on digital platforms as a four step process, visualised in Figure~\ref{fig:model-diagram}:
When a user wants to find a service provider, 1) the platform recommends them a list of possible choices, taking into account their rating and possibly their group.
Then, 2) the user chooses a provider in a boundedly rational way, 3) they subsequently interact, and then 4) the user truthfully reports how the interaction went back to the platform. This process then repeats whenever a user looks for a provider.

Given that the number of users on the platform is typically multiple orders of magnitude larger than the number of providers,\footnote{In 2020, when Airbnb most recently published their internal statistics, there were 14.1 million listings that had been booked in the last 365 days, with various sources reporting between 2.9 and 5 million hosts and 150 million users.} we assume a finite population size $Z$ of partners, $Z_D$ of these being from some dominant social group and $Z_M$ being marginalised, and assume for simplicity that the user population is infinitely large, as in \citet{monachou_discrimination_2019} and \citet{che_statistical_2024}.
This partial mean-field assumption, where one population is finite and the other infinite, allows us to neatly represent the user population using three parameters which express how users interact with different aspects of the interaction process previously described.

First, we use $\epsilon$ to measure the degree of \textbf{rating bias} experienced by marginalised users.
We assume that all providers have an individual rating which is seen by users as either ``good enough'' or ``not good enough'' which we refer to as \textbf{Good} ($G$) and \textbf{Bad} ($B$) respectively.
For simplicity, we assume that after every interaction, the involved user non-strategically reports the action taken by the provider to the platform.
This action can be either high-effort ($H$) or low-effort ($L$), which, when reported to the platform, is translated into the corresponding rating $G$ or $B$, overriding the provider's current reputation and reflecting the ``prosociality'' of their most recent action.
We introduce $\epsilon \in [0, 1]$ to be the likelihood that a provider of the marginalised group who plays $H$ is falsely reported to have played $L$, which we refer to as the \textbf{rating bias}.
We assume that, even if users do not know the group of their interaction partner before it takes place, they do know it by the time they write the review given that they have communicated and sometimes met online or in person.

The plausibility of modelling ratings and actions with only two options is derived from platforms' implementations of punishments for low ratings.
The practice of removing partners with ratings below a certain threshold is a policy of both Airbnb\footnote{https://www.airbnb.com/help/article/2895} and Uber\footnote{https://www.uber.com/us/en/drive/driver-app/deactivation-review/}, the latter even leading to a high-profile lawsuit alleging that this practice violated the Civil Right Act.
Furthermore, because ratings are often so heavily skewed towards the positive end, it can be difficult to meaningfully distinguish between ratings beyond whether they are above or below some platform-specific threshold, something reflected in folk-advice written by/for users\footnote{https://www.kenrockwell.com/tech/ebay/index.htm}.
Because of this, a single bad rating can be enough to put a provider below the platform's threshold for algorithmic preference and cause users to take caution before selecting the provider.

We measure how ``hard'' this rating threshold is, and in turn the amount of caution displayed by users, by $\gamma \in [0, 1]$, which we call the \textbf{rating sensitivity}.
This parameter can alternatively be thought of as the trust users have in the rating system.
When comparing between a list of providers, we assume that  the platform obfuscates the demographic characteristics of the providers so users can only decide between providers using their rating.
If $\gamma = 1$, then users will simply randomise amongst providers with rating $G$, but with $\gamma < 1$, users assign weight $1-\gamma$ to providers with rating $B$ and weight 1 to providers with rate $G$, and randomise amongst \textit{all} providers, selecting each with probability proportional to their weight.
If $\gamma=0$, users are indifferent between $G$ or $B$ providers.

Finally, we measure the number of providers that users will compare before making a selection with $k$, this is also known in business terms as \textbf{consumer involvement}.
It has long been known that users selections are \textit{heavily} skewed towards the first few results~\cite{keane_are_2008}, and while this may be controlled to some extent by the platform (e.g. Uber's fully automated matching), consumer involvement is more often explained by characteristics inherent to the market of which the platform is a part~\cite{niosi_involvement_2021}.
In the remainder of this paper, we refer to a user population by its rating bias, rating sensitivity, and consumer involvement written in a triple $(\epsilon, \gamma, k)$.

\subsection{Modelling the Recommender System}
To encourage users to engage in cooperative, high-effort behaviour, platforms use recommender systems to prioritise partners with high ratings, making users more likely to interact with those deemed ``high quality''.
Rather than describing any particular algorithm or recommender system, we look at the output produced by whichever system is chosen by the platform.
As providers who sign up to digital service platforms typically have to provide identification, we assume that platforms are able to infer the group of a provider.

Given some user population with consumer involvement $k$, we measure the extent of a platform's ``high rating prioritisation'' by integer $k_G \leq k$.
This value deterministically guarantees that at least $k_G$ of the search results are rated $G$.
Additionally, as some platform designers recognise that their ratings are biased and want to explicitly counteract this, we introduce $k_M \leq k_G$ which, again, guarantees that at least $k_M$ of the $k_G$ providers will be from the marginalised group.
To account for the fact that, depending on the number of providers playing each strategy, there may not always be $k_G$ or $k_M$ providers that fit the requirements to be prioritised, we let $Z_G$, $Z_{GM}$, and $Z_{GD}$ be the number of providers, marginalised providers, and dominant providers rated $G$ respectively and define $\hat{k}_G := \min(k_G, Z_G)$ and $\hat{k}_M := \min(k_M, Z_{GM})$.

Assuming that the remaining $k_R := k - \hat{k}_G$ providers are sampled from all providers \textit{after} the $\hat{k}_M$, then $\hat{k}_G - \hat{k}_M$ providers are sampled from their respective sub-populations, we can calculate the probability that a provider in group $g \in \{M, D\}$ with rating $r \in \{B, G\}$ is included in any particular list of search results and subsequently chosen given that they were included.
Because the likelihood that one is chosen depends on who else is included, and this in turn depends on at which stage ($\hat{k}_M$, $\hat{k}_G$, or $k_R$) one is chosen, we relegate the calculation of this probability to appendix Section~1 and subsequently refer to its value as $p(g,r \mid Z_{GD}, Z_{GM})$.


\subsection{Utility calculation}
In our model, the expected utility of a provider is their payoff from a single interaction multiplied by the likelihood that the provider is chosen for any given interaction.
A high-effort ($H$) provider engages in behaviours such as ensuring cleanliness of a property and communicating in a timely manner with users, which we assume has per-interaction a utility cost $c > 0$.
On the other hand, we assume that a low-effort ($L$) provider does not have to pay this cost, and that both types of provider have the same per-interaction revenue $b$ with $b > c > 0$.
If $c$ and $b$ are close, then margins in the market are tight and there is a large incentive to play low-effort ($L$).

As the likelihood that a partner is chosen depends on the strategies played by the rest of the population, our model's dynamics consider every agent's individual incentive to switch between their current strategy and the other, which we detail in Section~\ref{sec:model-strategy-dynamics}.
We write the state space of the dynamics as $\mathbf{h} = (h_M, h_D)$, where $h_i$ represents the number of agents playing $H$ in group $i$ (so $0 \leq h_i \leq Z_i$), see how $\mathbf{h}$ varies over time subject to stochastic dynamics, and then weigh each provider's utility in each state by the time the dynamics spend in that particular state.

At any point in time, all of the $h_M$ marginalised $H$-playing providers have a $1-\epsilon$ probability of correctly being assigned rating $G$ from their last interaction, and so $Z_{GM}\sim\textrm{Binomial}(h_M,1-\epsilon)$ with p.m.f $f_{Z_{GM}}$.
Therefore, to calculate the likelihood of a provider in group $g$ playing $s$ being chosen, we sum over their rating $r$ and over $Z_{GM}$ given $g$ and $r$ (as when $(g,r) = (M,G)$, one partner is already accounted for), calculating $p(g,r\mid Z_{GD}, Z_{GM})$ for each case as outlined in appendix Section~1.
$p(g,r\mid Z_{GD}, Z_{GM})$ represents the likelihood that a partner from group $g$ with rating $r$ is shown in the search results given the ratings of others, capturing competition among providers.
As such, the utility $u$ of a provider in group $g$, playing strategy $s\in \{L, H\}$ given the (competing) provider population is $\mathbf{h}$, is

\begin{align}\label{expr:expected-utility}
u(g,s\mid\mathbf{h}) &= \pi(s)\sum_{r\in\{B,G\}}p(r \mid (g, s)) \sum\limits_{z=0}^{h_M}f_{Z_{GM}}(z) p(g,r\mid Z_{GD}, z),
\end{align}

where $\pi(s) = (b-c\,\mathbb{I}_{\{s=H\}})$ and $\mathbb{I}$ is the indicator function, and


\begin{equation}
    p(G \mid (g, s)) = \begin{cases}
        0 ,&\textrm{if}\,s=L\\
        1-\epsilon,&\textrm{if}\,s=H\,\textrm{and}\,g=M\\
        1,&\textrm{if}\,s=H\,\textrm{and}\,g=D.
    \end{cases}
\end{equation}

\subsection{Population Dynamics}\label{sec:model-strategy-dynamics}

We model variations in $\textbf{h}_t = (h_D, h_M)_t$ over time using a frequency-dependent Moran process~\cite{moran_random_1958} as in other models studying the emergence of cooperation~\cite{nowak_emergence_2004, traulsen_stochastic_2006}.
A Moran process is a stochastic evolutionary model used in genetics where the size of each ``allele'' (in our case strategy) remains constant, and where alleles with a fitness advantage tend to become more prevalent in the population.

We apply this model as such:
each timestep $t$, we sample a random focal agent $i$ and flip their strategy $s_i$ (either from $H$ to $L$ or vice-versa) with probability $\mu$, and otherwise compare their utility $u(g_i,s_i)$ to another randomly sampled agent from their group $j$ using the \textbf{Fermi pairwise imitation rule}~\cite{traulsen_stochastic_2006}:
Let $\Delta u^g_{ij}(\mathbf{h}) := u(g,j\mid\mathbf{h})-u(g,i\mid\mathbf{h})$ using the definition of $u(g,s\mid\mathbf{h})$ in (\ref{expr:expected-utility}) and let $\beta \in \mathbb{R}_+$ be the strength of selection, the probability that the number of agents playing $H$ increases ($f^g_+$) or decreases ($f^g_-$) is given by

\begin{align}
    f^{g}_\pm (\mathbf{h}_t) = \frac{1}{1+e^{\mp\beta\Delta u^g_{HL}(\mathbf{h}_t)}}.
\end{align}

This process of mutation and replacement, allows us to define a discrete-time Markov chain $\mathbf{h}_t$ on the state space of the number of agents playing $H$ in each group
\[
\mathcal{S} = \{(h_M, h_D): 0 \leq h_M \leq Z_M, 0 \leq h_D \leq Z_D\},
\]
(ordered lexicographically), initial state $\mathbf{h}_0$, mutation probability $\mu$, and transition matrix $P^{\mathcal{S}\times\mathcal{S}}$ with entries
\begin{align}
&P_{\mathbf{h}_t,\mathbf{h}_{t+1}} = \\
&\begin{cases}
  \mu\frac{Z_D-h_D}{Z} + (1-\mu)\frac{Z_D-h_D}{Z_D}\frac{h_D}{Z_D-1} f^D_+(\mathbf{h}_t),&\:\textrm{if}\,\textbf{h}_{t+1} = (h_D + 1, h_M)\\[7pt] 
  \mu\frac{h_D}{Z} + (1-\mu)\frac{h_D}{Z_D}\frac{Z_D-h_D}{Z_D-1} f^D_-(\mathbf{h}_t),&\:\textrm{if}\,\textbf{h}_{t+1} = (h_D - 1, h_M)\\[7pt]
    \mu\frac{Z_M-h_M}{Z} + (1-\mu)\frac{Z_M-h_M}{Z_M}\frac{h_M}{Z_M-1} f^M_+(\mathbf{h}_t),&\:\textrm{if}\,\textbf{h}_{t+1} = (h_D, h_M+ 1)\\[7pt]
  \mu\frac{h_M}{Z} + (1-\mu)\frac{h_M}{Z_M}\frac{Z_M-h_M}{Z_M-1} f^M_-(\mathbf{h}_t),&\:\textrm{if}\,\textbf{h}_{t+1} = (h_D, h_M-1)\\[7pt]
  1-\sum\limits_{\mathbf{x}\neq \mathbf{h}_t} P_{\mathbf{h}_t,\mathbf{x}},&\:\textrm{if}\,\mathbf{h}_{t+1} = \mathbf{h}_t\\
  0,&\textrm{otherwise.}\\
\end{cases}\nonumber
\end{align}

The Markov chain $\mathbf{h}_t$ is irreducible if $\mu > 0$ as mutations draw the chain out of absorbing states, and aperiodic as the chain has a strictly non-zero probability of staying in the same place for all $\beta \neq \infty$.
Therefore, we can find its stationary distribution $\mathbf{h^\ast}$ defined as the limit of the recurrence relation $\mathbf{h}_{t+1} = \mathbf{h}_t P$ by solving the linear system $(I - P)^T \mathbf{h}^{\ast T} = \mathbf{0}$ subject to the constraint $\sum_i\mathbf{h}^{\ast}_i = 1$.

\section{Experimental Setup}\label{sec:experiments}
In our experiments we aim to explore how different user populations $(\epsilon, \gamma, k)$ and recommender system parameters $(k_G, k_M)$ affect 1) the incentive for each group to cooperate (play $H$), 2) the value users get out of the platform, and 3) the average utility of the two groups of providers. In Section~\ref{sec:evaluation-metrics} we give a precise definition to each of these metrics, and below we detail the parameters common to every experiment we run.

To minimise differences in utility that are incidental to (relative) group size we set the size of both groups to be equal: $Z_D = Z_M = 20$ and keep this fixed from here on, referring readers to appendix Section~2 for results with $Z=20$ and $Z=80$.
While this is certainly scaled back when compared to some real online platforms, it is large enough that $k_G$ and $k_M$ can take many values between 0 and $k \leq Z$, but also small enough such that parameter grid searches are computationally feasible (recall every simulation requires inverting a $(Z_D\,Z_M) \times (Z_D\,Z_M)$ matrix).
Results for other population sizes, larger and smaller, are qualitatively the same, \textit{mutatis mutandis}.
As $\mu$ and $\beta$ are both unitless parameters of evolution, we can scale one in terms of the other to achieve the same result.
As such, we arbitrarily set the mutation rate $\mu = 1/Z = 1/40$, such that, on average, once in every $Z$ strategic update steps the updating provider will randomise their strategy rather than imitate another provider.

To both emphasise the effect of platform design choices and simulate a case facing many service providers where profit margins are relatively tight, we set $c=1$ and $b=1.2$, raising $b$ or lowering $c$ simply makes cooperation easier to sufficiently incentivise.
Given that our utilities are calculated as the likelihood of an interaction multiplied by the payoff from an interaction, if selections were done completely at random then each agent would have an average utility proportional to $1/40$.
The selection strength $\beta$ is set out of modelling convenience such that $\beta\times\Delta u$ has an order of magnitude close to $1$, attempting keeping the output of the Fermi function away from its region of near-zero gradient, leading to our choice of $\beta = Z/(b-c) = 20$.

We initially hold $k_M = 0$, which represents the status-quo where, for implementation or political reasons, platforms do not explicitly prioritise marginalised providers.
After exploring the dynamics of the model in Section~\ref{sec:res-strategy-dynamics} and introducing the ``user-provider trade-off'' subject to this restriction in Section~\ref{sec:pareto-front}, we allow platform designers to alter $k_M$ while keeping $k_G$ fixed, simulating a situation where platform designers want to minimally alter the user experience in Section~\ref{sec:anti-discrimination}.
Finally, we remove all restrictions on $k_G$ and $k_M$ and introduce uncertainty on the precise value of $\epsilon$, seeing how increasing uncertainty affects demographic parity in Section~\ref{sec:uncertainty}.

\subsection{Evaluation Metrics}\label{sec:evaluation-metrics}
\paragraph{Mostly cooperative}
A single group $g \in \{M, D\}$ is \textbf{mostly cooperative} if the stationary distribution of the strategy dynamics $\mathbf{h}^\ast$ spends most of its time on the edge $h_g = Z_g$ i.e. the sum of $\mathbf{h}^\ast$ over all states with $h_g = Z_g$ is greater than 0.5.
A platform is mostly cooperative if both of its groups are.

\paragraph{User Experience}
Given the stationary distribution $\mathbf{h}^\ast$, the \textbf{user experience (UX)} on the platform is defined as the likelihood that the service provider plays $H$.
Let $S$ be the state matrix whose entries are the number of agents playing $H$ in each group
    \[
    S  = (S_1, S_2) := \begin{bmatrix}
    (0, 0) & (0, 1) &\cdots & (0, Z_D)\\
    (1, 0) & (1, 1) & \cdots & (1, Z_D)\\
    \vdots &  \vdots & \ddots & \vdots\\
    (Z_M, 0) & (Z_M, 1) &\cdots & (Z_M, Z_D)
\end{bmatrix}
\]
then define $\hat{S}_M := \frac{S_1}{Z_M}$, and $\hat{S}_D :=\frac{S_2}{Z_D}$.
Then, we can calculate the expected strategy in each group ($\sigma^\ast$) by taking the elementwise product ($\odot$) of $\hat{S}$ with $\mathbf{h}^\ast$, before summing over the rows and columns, which is equivalent to taking the matrix $1$-norm as $S$ is non-negative:
\begin{equation}
    \sigma^\ast = \| \hat{S} \odot \mathbf{h}^\ast\|_1
\end{equation}
Finally, we can calculate
\begin{equation}
    \mathbf{UX} = \frac{\sigma^\ast_M Z_M + \sigma^\ast_D Z_D}{Z_M+Z_D}
\end{equation}

\paragraph{Demographic Parity Ratio}
As we discuss in Section~\ref{sec:related-lit}, we measure fairness, for simplicity, through the \textbf{demographic parity ratio (DPR)}, which captures the ratio between the average utility of the better-off and worse-off group of service providers.

Define matrix $(U^g)_{\substack{0 \leq i \leq Z_M\\0 \leq j \leq Z_D}}$ with entries
\[U^g_{i,j} := u(g, L\mid\mathbf{h}=(i,j))\frac{Z_g-h_g}{Z_g} + u(g, H\mid\mathbf{h}=(i,j))\frac{h_g}{Z_g}\]
to be the average utility in group $g$ at each state where $h_g$ is the number of $H$-players in group $g$, taking either value $i$ or $j$ depending on which group is being considered.
Then we can calculate
\begin{equation}
    \textbf{DPR} := \frac{\min\left(\|\mathbf{h}_M^\ast \odot U^M\|_1, \|\mathbf{h}_D^\ast \odot U^D\|_1\right)}{\max\left(\|\mathbf{h}_M^\ast \odot U^M\|_1, \|\mathbf{h}_D^\ast \odot U^D\|_1 \right)}
\end{equation}

\paragraph{Pareto Optimality}
Given user population $(\epsilon, \gamma,k)$, a platform designer's choice of $(k_G, k_M)$, potentially subject to the constraint $k_M=0$, is \textbf{Pareto optimal} if a different choice could not simultaneously increase the platform's \textbf{UX} and \textbf{DPR}.
The set of choices of $(k_G, k_M)$ that are Pareto optimal is called the \textbf{Pareto front}.

\section{Results}
\subsection{Achieving a cooperative baseline}\label{sec:res-strategy-dynamics}
\begin{figure*}[tb]
    \centering\includegraphics[scale=0.65]{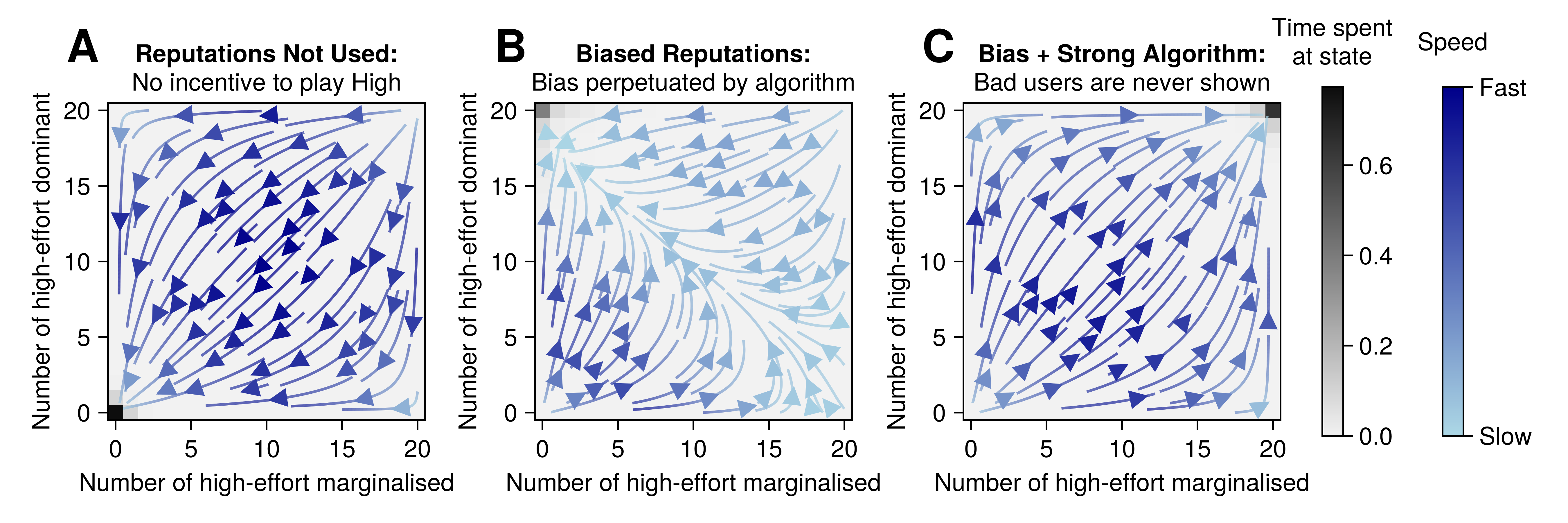}
    \caption{
        We demonstrate how changing platform design affects strategy dynamics for providers.
        The proportion of time the dynamics spend at each state under the stationary distribution is plotted as a greyscale heatmap, and the average direction of the dynamics are overlaid as a stream plot.
        All subplots have $k=10$ and $\epsilon=0.3$.
        For subplot~\textbf{A}, we set $k_G = \gamma = 0$, which means that $\epsilon$ could take any value without affecting the dynamics.
        Subplot~\textbf{B} and \textbf{C} have $k_G=5$ and $k_G=10 = k$ respectively. We observe that the platform's design choices impact the providers' strategic dynamics, leading them to adopt low-effort strategies (A), exacerbating biases by eliciting high-effort only from the dominant group (B) or inspiring high-effort by all groups (C).
    }
    \label{fig:dynamics_over_heatmap}
\end{figure*}

At a fundamental level, every platform must maintain an incentive for providers to be high-effort in order to ensure users have a good experience, given user parameters $(\epsilon, \gamma, k)$.
In Figure~\ref{fig:dynamics_over_heatmap}, we demonstrate the varying strategy dynamics that respective populations of providers have in three scenarios with varying user parameters and $k_G$, and who all have $k_M = 0$.

The left-most scenario (\textbf{A}) effectively has no reputation system ($k_G = \gamma = 0)$, meaning neither the search algorithm nor the users use any information about the quality of a potential provider in their decision making.
Unsurprisingly, without an incentive to be high-effort, no providers choose to do so.
This outcome is characteristic of any platform that, given a certain user population, fails to sufficiently incentivise providers to cooperate.

The centre and right-most platform with $k=10$, \textbf{B} and \textbf{C}, show how user populations with non-zero rating bias (here $\epsilon = 0.3$) and/or low sensitivity to ratings (here $\gamma=0.6$) can lead to either a separating or pooling equilibrium depending on the platform's choice of $k_G$.
In the former case ($k_G = 5$), although the dominant group is ``mostly cooperative'' as previously defined in Section~\ref{sec:evaluation-metrics}, the incentives are not sufficient for marginalised users to play $H$ (high-effort) which we speculate is due to the difficulty maintaining a high enough rating to be prioritised by the recommender system and user population, and the insufficient prioritisation when their rating is high.
This leads to the divergence of strategy between dominant and marginalised providers that is not found in platform \textbf{C} ($k_G = 10 = k$), a platform that \emph{does} feature sufficient incentives for both groups to be mostly cooperative.


\subsection{The $(k_G, k_M=0)$ Pareto front}\label{sec:pareto-front}
\begin{figure}[tb]
    \centering\includegraphics[scale=0.65]{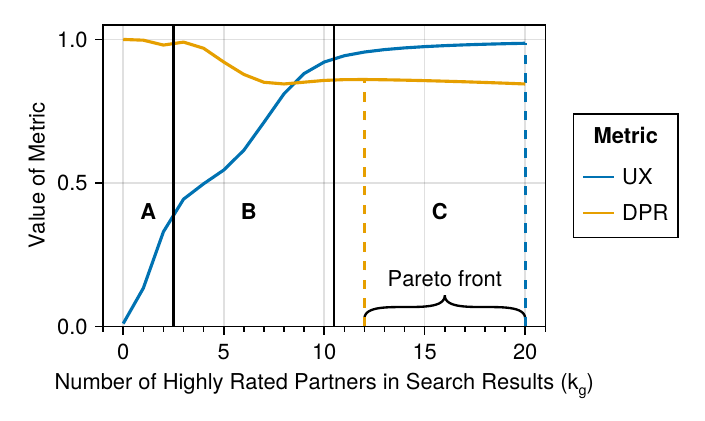}
    \caption{
        In this Figure we fix the user population to have rating bias $\epsilon=0.15$ and rating sensitivity $\gamma=0.8$, and vary $k_G$ while holding $k=20$.
        We can partition the x-axis into three distinct regions defined by whether no groups, only the dominant group, or both groups are ``mostly cooperative''.
        We indicate the maximum value attained by each of the three evaluation metrics defined in Section~\ref{sec:evaluation-metrics} using a dashed line.
    }
    \label{fig:platform-parameters-lines}
\end{figure}

In Figure~\ref{fig:platform-parameters-lines}, we show the \textbf{UX} and \textbf{DPR} of a platform with population parameters $k=20$, $\epsilon=0.15$, and $\gamma=0.8$ and $k_M=0$ subject to different choices of $k_G$.
The dynamics of the three platforms \textbf{A}, \textbf{B}, and \textbf{C} from Figure~\ref{fig:dynamics_over_heatmap} represent three qualitatively different regimes of cooperation (or lack thereof) which can be induced through choice of $k_G$ and $k_M$.
The size of these regimes in ($k_G, k_M$)-space depends on the population parameters.
As in this case $k_M=0$, we plot the $k_G$-changepoints of the regimes as black vertical lines and label the regions correspondingly, also marking the maximum values attained by each metric in regime \textbf{C}.

Values of $k_G$ between these maximum values make up Pareto front.
Define $k_G^{\textbf{UX}}$ and $k_G^{\textbf{DPR}}$ to be the values of $k_G$ maximise \textbf{UX} and \textbf{DPR} respectively while inducing regime \textbf{C} dynamics.
When $k_M = 0$ we have $k^{\textbf{UX}}_G \equiv k$ as \textbf{UX} is always monotonically increasing in $k_G$:
the more highly-rated providers that users are shown, the more likely they are to choose a provider playing $H$.\footnote{This would still be true if there were a chance for $L$-players to be $G$-rated as long as we are in regime \textbf{C}. In this case, no matter the false positive rate, because more agents are cooperating than defecting, showing more $G$-rated agents still increases the likelihood that a user selects an $H$-player}
However, for $k_G > k^{\textbf{DPR}}_G$, the increased pressure to cooperate is sufficiently small that the emergent effect is just showing marginalised providers less often due to the rating bias.
This subsequently decreases the share of interactions they are involved in and lowers the \textbf{DPR}.

\begin{figure}[tb]
    \centering\includegraphics[scale=0.65]{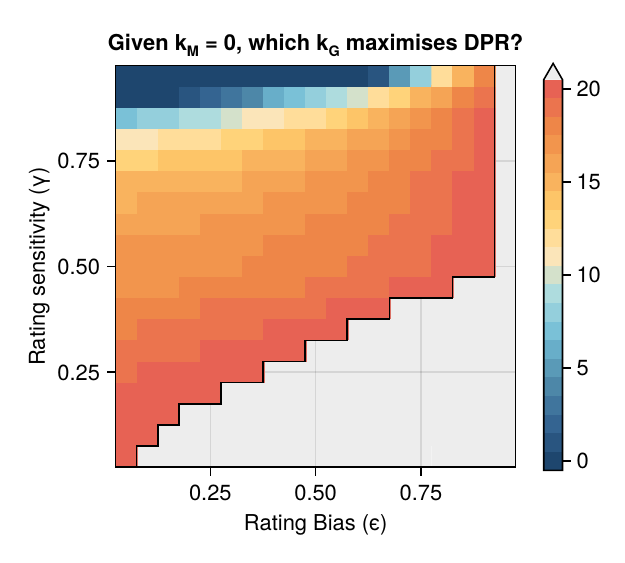}
    \caption{
        For $k=20$ (as in Figure~\ref{fig:platform-parameters-lines}), we vary $\epsilon$ and $\gamma$, calculating the value of $k_G$ that  maximises the demographic parity ratio (\textbf{DPR}) while leaving the dynamics in regime \textbf{C}.
    }
    \label{fig:compare-ps-dp}
\end{figure}

The decreasing nature of \textbf{DPR} for $k_G > k_G^\textbf{DPR}$, combined with \textbf{UX} being monotonically increasing in the same region, implies that when $k_M = 0$, the Pareto front is the interval $[k_G^\textbf{DPR}, k]$.
In Figure~\ref{fig:compare-ps-dp}, we show the size of the Pareto front when $k=20$ (as in Figure~\ref{fig:platform-parameters-lines}), varying $\epsilon$ and $\gamma$ in the open set $(0, 1)$.
The reason we exclude the edges is that one of the metrics becomes completely flat (e.g. $\epsilon = 0 \implies\textbf{DPR} \equiv 1$), but floating point error causes distracting fluctuations.
We note that $k_G^\textbf{DPR}$ monotonically increases with $\epsilon$ and $\gamma$, meaning biased and unreliable reputations require that platforms put more effort into incentivising cooperation, particularly in the marginalised group, to achieve maximum demographic parity.





This analysis shows that if platforms are unwilling or unable to introduce active countermeasures against inherent rating bias~($\epsilon$) at the algorithmic level (i.e. $k_M=0$), then they have to choose between a better user experience and a more equitable provider experience, where, even given optimal choices, one necessarily comes at the cost of the other.

\subsection{Explicit anti-discrimination with $k_M \geq 0$}\label{sec:anti-discrimination}
\begin{figure}[tb]
    \centering\includegraphics[scale=0.65]{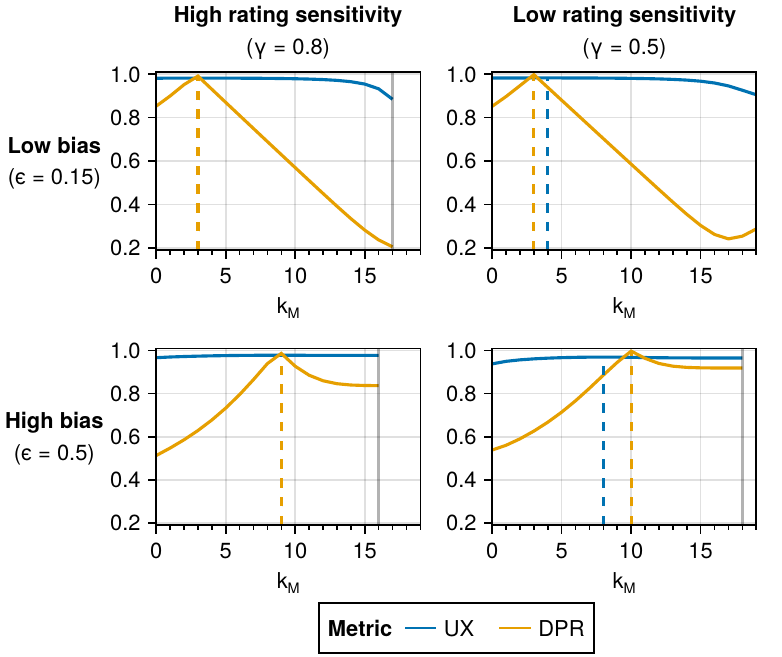}
    \caption{
        Consider a platform with $k=20$ and vary $\epsilon$ and $\gamma$ between low and high values.
        For each resulting scenario, find the value of $k_G$ that maximises $\mathbf{UX} \times \mathbf{DPR}$ subject to $k_M = 0$, we plot this as a faint grey vertical line.
        As we now vary $k_M$, the dynamics stay inside regime \textbf{C} and the evaluation metrics peak at the same value of $k_M$ when $\gamma$ is high enough (0.8).
    }
    \label{fig:k_M-platform-parameters-lines}
\end{figure}

We first assume that the platform has a fixed $k_G$, which could be for many reasons including wanting to maintain the overall ``feel'' of the platform to users by showing the same variety of ratings, and can choose any $k_M \geq 0$.
In Figure~\ref{fig:k_M-platform-parameters-lines}, we plot four different platforms.
For each we arbitrarily choose $k_G$ that maximises $\textbf{UX}\times\textbf{DPR}$, a value somewhere in the Pareto front, subject to $k_M = 0$, emulating the situation where a platform that cares about \textbf{DPR} wants to introduce an explicitly biased algorithm while making minimal changes to the platform from a user point of view.

We find that the higher the rating bias, the higher $k_M$ is required to be to correct for it: when $\gamma=0.8$, raising $\epsilon$ from 0.15 to 0.5 shifts the optimal value of $k_M$ from 3 to 9.
While changing $k_M$ has a large effect on fairness, it has almost no effect on users, evidenced by the relatively flatness of the \textbf{UX} line compared to \textbf{DPR}.
This is because raising $k_M$ does not change the number of high-effort agents shown to users, simply the demographics of the agents shown.
Only when $\epsilon$ is low and $k_M$ is very high do we see a fall in \textbf{UX}.
In this case the incentive for the dominant group to play $H$ is severely impacted due to them almost exclusively being shown only as one of the providers in the random sampling stage which makes up only a small fraction $k_R/k$ of all providers shown.
Because of this, they are shown almost as often when playing $L$ as when playing $H$.

While these results are promising, as any variable not controlled by the platform must be inferred from data, the designers may not have accurate values of $k$, $\epsilon$, and $\gamma$.
The sharp peak of \textbf{DPR} in Figure~\ref{fig:k_M-platform-parameters-lines} with respect to $k_M$ raises an issue for platform designers: by altering $k_G$ and $k_M$ you risk over- or under-correcting for bias.
As inaction is less harshly judged than poor action, it is important to see that a $k_M > 0$ policy is still effective under uncertainty.

\subsection{Optimising with uncertainty in rating bias}\label{sec:uncertainty}
\begin{figure}[tb]
    \centering\includegraphics[scale=0.65]{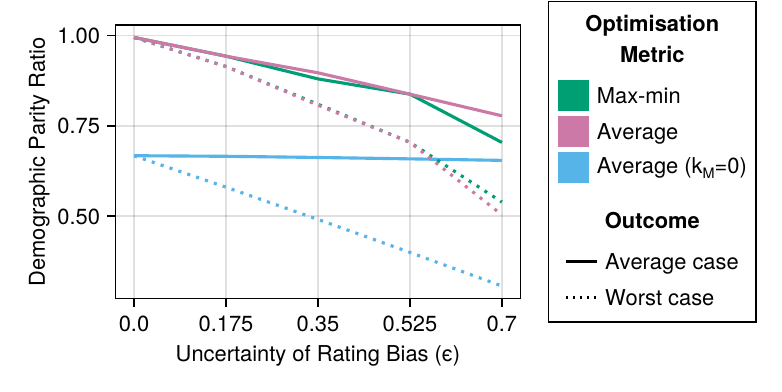}
    \caption{Let k = $20$, $\epsilon = 0.35$, and $\gamma=0.65$, and vary the uncertainty of rating bias. As previously established: forcing $k_M = 0$ leads to a low \textbf{DPR}. Even when uncertainty is 0.7, allowing $k_M \geq 0$ still results in a better worst-case outcome. Optimisation of max-min \textbf{DPR} and average \textbf{DPR} are both fine choices of metric to optimise over that only diverge significantly in outcome for very large uncertainty.
    }
    \label{fig:uncertainty-epsilon}
\end{figure}

Suppose the true, unobserved amount of rating bias exhibited by a population of users was $\epsilon = 0.35$, and let $\gamma = 0.8$ and $k = 7$ be observed.
For simplicity, we assume that the unobserved value of $\epsilon$ can be estimated to be within some interval $[\epsilon_\textrm{min}, \epsilon_\textrm{max}]$ with uniform likelihood over the interval.
We refer to the width of this interval ($\epsilon_\textrm{max}-\epsilon_\textrm{min}$) as the \textbf{uncertainty} in $\epsilon$.

Given some level of uncertainty, which values of $k_G$ and $k_M$ should the platform choose?
One might expect this to depend considerably on what the platform tries to optimise for.
Of course, the choice of values must leave the dynamics in regime \textbf{C}, but should you try to optimise the expected \textbf{DPR} or conservatively try to maximise its minimum value?
Surprisingly, these two straightforward metrics tend to yield very similar outcomes for low to moderate values of uncertainty, diverging only when uncertainty is very large.
This can be seen in Figure~\ref{fig:uncertainty-epsilon}, where until the uncertainty is 0.525, representing a range of $[0.0875, 0.6125]$, both the average- and worst-case outcomes of applying either a maximin strategy or maximising the expected value are almost identical.
After this point, the two diverge but importantly both still maintain better performance than a platform which keeps $k_M=0$.
In other words, even when uncertainty is very large, optimising for the worst-case outcome over $k_M \geq 0$ achieves better worst- and even \textit{average}-case performance than leaving $k_M = 0$, highlighting how risk-free it is to allow $k_M$ to vary.\balance

\section{Conclusion}\label{sec:conclusion}
Online marketplaces have become ubiquitous in modern society.
In this paper we develop a model that shows how algorithmic design decisions, applied in such online platforms, affect utility and fairness of users and providers. 
Our findings reveal that recommender systems can play a decisive role in undoing the effects of rating bias, a phenomenon that is pervasive across platforms.
However, if considerations of fairness take a back seat to the experience of the user population, then we show how seemingly fairness-neutral decisions can counterintuitively lead to less fair outcomes.
The low barrier of entry to employment provided by these platforms means they are relied upon by many economically and legally vulnerable members of society who themselves are often part of the groups that are discriminated against.
As such, designing these platforms with the goal of fair treatment of marginalised communities despite the inherent biases of users is a significant step towards fairness in the labour market more broadly.

By keeping the recommender system, user population, and the interaction itself abstract, we hope that this model might inform the application of complex adaptive systems, particularly evolutionary game theory, to more elaborate, tailored models in (digital) labour economics and industrial organisation:
while we assume a the common assumption of a constant number of providers, in real markets participants are constantly arriving, and existing agents may leave during downturns in their utility which is modelled in~\cite{monachou_discrimination_2019}.
Furthermore, our mean-field assumption on users and use of indirect as opposed to direct reciprocity may not be valid for smaller markets where consumers and producers are more equally numbered.
We ignore providers' forms of collective action, which might also affect fairness dynamics \cite{pilatti2024systematic,sigg2025decline}, a topic of natural interest to be explored in future extensions of our model. 
Finally, we assume that providers' demographic information is fully obfuscated, but this is obviously idealised and not typically the case~\cite{kricheli-katz_how_2016}.





Despite these limitations and suggestions for future work, our work already stresses the positive outcomes of anti-discrimination policies in digital economy platforms, and we test concrete interventions to improve fairness in systems where ratings are used to promote cooperative behaviour.



\begin{acks}
FPS acknowledges funding through ERC grant (RE-LINK,\\  https://doi.org/10.3030/101116987). 
\end{acks}



\bibliographystyle{ACM-Reference-Format} 
\bibliography{references}


\end{document}